\documentclass[useAMS,usenatbib,twocolumn,preprint]{mn2e}
\usepackage{amsmath}
\usepackage{graphicx}
\usepackage{oldlfont}
\usepackage{longtable}
\usepackage{rotating}
\usepackage{amssymb}
\usepackage[hang]{subfigure}
\def\jnl@style{\it}
\def\aaref@jnl#1{{\jnl@style#1}}
\def\aaref@jnl#1{{\jnl@style#1}}

\def\aj{\aaref@jnl{AJ}}                   % Astronomical Journal
\def\apj{\aaref@jnl{ApJ}}                 % Astrophysical Journal
\def\apjl{\aaref@jnl{ApJ}}                % Astrophysical Journal, Letters
\def\apjs{\aaref@jnl{ApJS}}               % Astrophysical Journal, Supplement
\def\apss{\aaref@jnl{Ap\&SS}}             % Astrophysics and Space Science
\def\aap{\aaref@jnl{A\&A}}                % Astronomy and Astrophysics
\def\aapr{\aaref@jnl{A\&A~Rev.}}          % Astronomy and Astrophysics Reviews
\def\aaps{\aaref@jnl{A\&AS}}              % Astronomy and Astrophysics, Supplement
\def\mnras{\aaref@jnl{MNRAS}}             % Monthly Notices of the RAS
\def\prc{\aaref@jnl{Phys.~Rev.~C}}       % Physical Review C
\def\prd{\aaref@jnl{Phys.~Rev.~D}}        % Physical Review D
\def\prl{\aaref@jnl{Phys.~Rev.~Lett.}}    % Physical Review Letters
\def\qjras{\aaref@jnl{QJRAS}}             % Quarterly Journal of the RAS
\def\skytel{\aaref@jnl{S\&T}}             % Sky and Telescope
\def\ssr{\aaref@jnl{Space~Sci.~Rev.}}     % Space Science Reviews
\def\zap{\aaref@jnl{ZAp}}                 % Zeitschrift fuer Astrophysik
\def\nat{\aaref@jnl{Nature}}              % Nature
\def\aplett{\aaref@jnl{Astrophys.~Lett.}} % Astrophysics Letters
\def\apspr{\aaref@jnl{Astrophys.~Space~Phys.~Res.}} % Astrophysics Space Physics Research
\def\physrep{\aaref@jnl{Phys.~Rep.}}      % Physics Reports
\def\physscr{\aaref@jnl{Phys.~Scr}}       % Physica Scripta
\def\commat{\aaref@jnl{Comm.~Math.~Phys.}}		% Communications in Mathematical Physics
\def\science{\aaref@jnl{Science}}		% Science
\def\cqg{\aaref@jnl{Classical Quant.~Grav.}}		% Classical and Quantum Gravity
\def\jpcs{\aaref@jnl{JPCS}}					% Journal of Physics Conference Series
\def\ijmpd{\aaref@jnl{Int.~J.~Mod.~Phys.~D}}			% International Journal of Modern Physics D
\def\grg{\aaref@jnl{Gen.~Relat.~Gravit.}}		% General Relativity and Gravitation
\def\rpp{\aaref@jnl{Rep.~Prog.~Phys.}}		% Reports on Progress in Physics

\newcommand{\beq}{\begin{equation}}
\newcommand{\eeq}{\end{equation}}
\newcommand{\beqar}{\begin{eqnarray}}
\newcommand{\eeqar}{\end{eqnarray}}

\title{Magnetars oscillations in the presence of a crust } 

\author[ A. Colaiuda and K. D. Kokkotas]
{ A.Colaiuda$^1$\thanks{E-mail:colaiuda@tat.physik.uni-tuebingen.de},
and K. D. Kokkotas$^{1,2}$\thanks{E-mail:kostas.kokkotas@uni-tuebingen.de}
\\
  $^1$Theoretical Astrophysics, University of T\"{u}bingen,  IAAT, Auf der Morgenstelle 10, Tuebingen 72076, Germany\\
 $^2$Department of Physics, Aristotle University of Thessaloniki,
  Thessaloniki 54124, Greece \\}
\begin{document}
\maketitle
\begin{abstract}
We study axisymmetric perturbations of neutron star endowed with a strong magnetic field (magnetars), considering the coupled oscillations of the 
fluid core with the solid crust. We recover discrete oscillations based mainly  in the crust and a continuum in the core. We also
confirm the presence of ``discrete Alfv\'en  modes'' in the gap between two contiguous continua (see \cite{2011MNRAS.410.1036H}) and, in addition, we  can resolve some of them also inside the continua. Our results
can explain both the lower and the higher observed quasi periodical oscillations (QPOs) in SGR 1806-20 and
SGR 1900+14 and put constrains on the mass, radius and crust thickness of the two magnetars.
\end{abstract}

%%%%%%%%%%%%%%
\section{Introduction}
%%%%%%%%%%%%%%

Over the last few years, a number of observational discoveries have  brought
magnetars (ultra-magnetized isolated neutron stars) to the forefront of researchers attention. These extreme objects comprise the Anomalous X-ray Pulsars (AXPs; 10 objects) and the Soft Gamma-ray Repeaters (SGRs; 5
objects), which are observationally very similar classes in many respects.
They are all slowly rotating  X-ray pulsars with spin periods clustered in a narrow range ($P\sim$ 2--12\,s),
relatively large period derivatives ($\dot P \sim 10^{-13}-10^{-10}$s\,s$^{-1}$),
spin-down ages of $10^3-10^4$\,yr, and magnetic fields, as inferred from the classical magnetic dipole
spin-down formula, of $10^{14}-10^{15}$\,G  (for a recent review see \cite{2008A&ARv..15..225M}).

SGRs undergo periods of activity during which recurrent bursts with sub-second  duration and peak luminosities of $\sim 10^{38}-10^{41}$ erg/s are observed. SGRs also show, on rare occasions,  extreme events known as giant flares. These are characterized by an initial spike of duration comparable to that of recurrent bursts, but many orders of magnitude higher luminosity. Only three giant flares have so far been observed in over 30 yr of monitoring.

According to the magnetar model \citep{1993ApJ...408..194T,2001ApJ...561..980T}  energy is fed impulsively to the neutron star magnetosphere when local ``crustquakes'' let magnetic helicity propagate outwards, giving rise to recurrent bursts with a large range of amplitudes. Giant flares are believed to originate from large-scale rearrangements of the inner field or catastrophic instabilities in the magnetosphere \citep{2001ApJ...561..980T,2003MNRAS.346..540L}. Most of this energy breaks out of the magnetosphere in a fireball of plasma expanding at relativistic speeds which results in the initial spike of giant flares. The decaying, oscillating tail that follows the spike displays many tens of cycles at the neutron star spin rate. This is interpreted as being due to a ``trapped fireball'' which remains anchored inside the magnetosphere and cools down in a few minutes. The total energy released in this tail is $\sim 10^{44}$ erg in all three events detected so far.

%Until 2004, the energy budget of SGRs and AXPs was believed to be dominated by their persistent emission at a level of $\sim 10^{35}$erg\,s$^{-1}$. This translated into an internal field of $\sim 10^{15}$ G. The properties of the 2004 Dec 27 giant flare from SGR1806-20 imply that the emission budget of Magnetars is dominated by giant flares, see e.g. \cite{2009MmSAI..80..186S}. This therefore has important implications for several subjects at the forefront of research.

A power spectrum analysis of the high time resolution data from the 2004 Dec 27 event of SGR1806-20, observed with the X-Ray Timing Explorer (RXTE), led to the discovery of fast Quasi Periodic Oscillations (QPOs) in the X-ray flux of the decaying tail of SGR (\cite{I2005}). QPOs with different frequencies were detected, some of which were active simultaneously and displayed highly significant QPO signals at about 18, 26, 30, 93, 150, 625 and 1840 Hz \citep{WS2006a}. A re-analysis of the decaying tail data from the 1998 giant flare of another magnetar, SGR 1900+14, revealed QPOs around frequencies of 28, 54, 84 and 155 Hz \citep{SW2006}. Hints for a signal at $\sim$ 43 Hz in the March 1979 event from SGR 0526-66 were reported as early as 1983 \citep{1983A&A...126..400B}. All QPO signals show large amplitude variations with time and especially with the phase of the stars rotational modulation.

QPOs have also been argued to provide independent evidence for superstrong magnetic fields in SGRs
\cite{2007ApJ...661.1089V}. 
Numerous explanations have been proposed for the origin of the QPOs including the torsional oscillations of the crust alone  or even as global seismic vibration modes of magnetars \citep{2005ApJ...634L.153P,2007MNRAS.375..261S,SA2007,2009PhRvL.103r1101S}. Moreover,  \cite{2006MNRAS.368L..35L} argued that the QPOs may be driven by the global mode of the  magneto-hydrodynamic (MHD) fluid core of the neutron star and its crust, rather than the mechanical mode of the crust. Following this idea,  \cite{2008MNRAS.385L...5S,2009MNRAS.396.1441C,2009MNRAS.397.1607C} recently made two-dimensional numerical simulations (both linear and non-linear) and found that the  Alfv\'en oscillations form  continua which  may explain  the observed QPOs. The weak point of these simulations is the absence of the crust. Moreover, some of the expected eigenfrequencies \citep{1998ApJ...498L..45D,I2005,2005ApJ...634L.153P,2007MNRAS.375..261S,2008MNRAS.385L...5S,2009MNRAS.396.1441C,2009MNRAS.397.1607C} match the observed QPO frequencies found by \cite{SW2006}. These findings have opened the field of magnetar seismology, which   provides an unprecedented probe of the star's crust and interior. In retrospective, predictions made in \cite{2009MNRAS.396.1441C} led to re-analysis of the 2004 Dec 27 event and revealed some new frequencies \citep{2011A&A...528A..45H}.

In this work we move to a more general model for neutron star i.e. we assume a fluid interior and a solid crust which are threatened by a dipole magnetic field. In the interior the perturbations will be dominated by Alfv\'en waves propagating along the field lines forming a continuum spectrum  as it has been seen in  \citep{2008MNRAS.385L...5S,2009MNRAS.396.1441C,2009MNRAS.397.1607C} while in the crust torsional oscillations dominate \citep{2007MNRAS.375..261S}. 
Our results differ partially from the ones by \cite{2011MNRAS.410L..37G}, where the authors study a magnetised neutron stars with the presence of an elastic crust: in fact, we discover the presence of a type of ``discrete Alfv\'en modes'' which are not present in the  absence of a crust.  These modes have been seen  by \cite{2011MNRAS.410.1036H} in the gaps between two contiguous continua (for this reason they call them  also ``gaps modes"). However, we are  able to resolve these modes also inside the continua. 

%%%%%%%%%%%%%%%%
\section{Description of the problem}
%%%%%%%%%%%%%%%%

Here we give a brief description of the equations and the boundary conditions that we used in our study.
First, we consider a spherically symmetric and static star, described by the
TOV equations and the line element
%%%
\begin{equation}\label{ds}
ds^2=-e^{2\Phi(r)}dt^2+e^{2\Lambda(r)}dr^2+r^2(d\theta^2+\sin^2 \theta d\phi^2).
\end{equation}
%%%%%
Although here we used an ultra strong magnetic field $B=4 \times 10^{15}$ Gauss, we  neglect
the induced deformation due to magnetic pressure, since it is still small because the magnetic energy  
$E_m$ is a few orders of magnitude
smaller than the gravitational binding energy $E_g$. Typically, $E_m/E_g
\simeq 10^{-4}(B/10^{16}G)^2$,  see  \cite{2008MNRAS.385.2080C} and \cite{2008MNRAS.385..531H} for results  on magnetar's deformation.
Axial and polar perturbations don't couple to each other and could be evolved independently, because we consider pure axisymmetric perturbations on a spherically symmetric star. Note that this assumption is not valid in a non-axisymmetric background: or example, as we mentioned before, a strong magnetic field will slightly deform the star away from sphericity and in this case there is  a coupling of axial and polar oscillations.  We have chosen to ignore the effects of smaller deformations since although they will introduce a tiny coupling between the polar and axial modes, in practice the properties of the two spectra will not be affected neither qualitative or quantitative. Other effects, such as resistivity, can potentially have more significant contributions (see \cite{2010MNRAS.tmp.1884L}).

 In the following we  consider only torsional oscillations, which are of axial type and do not induce density variations in spherical 
stars: in this way, the radiative part of the metric describing the gravitational field does not vary significantly. For this reason the frequency of torsional 
oscillations is determined with satisfactory accuracy even when the metric perturbations were completely ignored by setting $\delta g_{\mu \nu} = 0$, i.e. working in the Cowling approximation (\cite{1941MNRAS.101..367C}).
In \cite{2007MNRAS.375..261S} the MHD oscillations of the equilibrium model were derived by transforming  
the perturbed linearized equation of motion and the perturbed magnetic induction equations into $2D$-wave equation for the displacement function
${\cal Y}(t,r,\theta)$ which is related to the contravariant component of the perturbed 4-velocity, $\delta u^\phi$ via
\begin{equation}
\delta u^\phi = e^{-\Phi} \partial_t {\cal Y} (t,r,\theta) \, ;
\end{equation}
see \cite{2007MNRAS.375..261S} for an analytical derivation .  
The $2D$-wave equation for the displacement 
${\cal Y}(t,r,\theta)$ has the following form:
\begin{equation}\label{2d}
A_{tt}\frac{\partial^2 {\cal Y}}{\partial t^2}=A_{20}\frac{\partial^2 {\cal Y}}{\partial r^2}
+A_{11}\frac{\partial^2 {\cal Y}}{\partial r \partial \theta}+A_{02}\frac{\partial^2 {\cal Y}}{\partial \theta ^2}
+A_{10}\frac{\partial {\cal Y}}{\partial r}+A_{01}\frac{\partial {\cal Y}}{\partial \theta}.
\end{equation}
All  coefficients $A_{tt}, A_{20},A_{11},A_{02},A_{10}$ and $A_{01}$ depend 
only on  $r$ and $\theta$, see \cite{2008MNRAS.385L...5S}. In \cite{2009MNRAS.396.1441C} % for the explicit form of all the coefficients.
  a coordinate transformation of the form
%%%%
\begin{equation}\label{new_trasf}
X=\pm\sqrt{a_1} \sin \theta  \quad \mbox{and} \quad  Y=\pm\sqrt{a_1} \cos \theta \, \, ,
\end{equation}
%%%
is used to transform the $2D$-wave equations into a $1D$-wave equation in  the part of the star where the shear modulus was zero (core). The final form of this equation is:
\begin{equation}\label{1d_n}
 A_{tt}\frac{\partial ^2 {\cal Y}}{\partial t^2}=\tilde{A}_{02}\frac{\partial^2 {\cal Y}}{\partial Y^2}+\tilde{A}_{01} \frac{\partial {\cal Y}}{\partial Y} \, \, ,
\end{equation}
where
\begin{eqnarray}\label{coeff_1d}
\tilde{A}_{02}&=&\frac{1}{4 \pi r^4}a_1 {a_1}'^{2}, \\ 
 \tilde{A}_{01}&=&\frac{X}{2\pi r^4}a_1\left(\frac{2}{r^2}a_1 e^{2\Lambda}-4\pi j_1e^{2\Lambda}-\frac{{a_1}'^{2}}{2}\right) \, \, .
\end{eqnarray}
Here $a_1(r)$ and $j_1(r)$ are the radial components of the electromagnetic 4-potential 
and  the 4-current and a prime indicates  derivative with respect  the radial coordinate,  respectively. Note that the function $a_1(r)$ is dimensionless. 

The distribution of the magnetic field inside the star can be found by solving the Grad-Shavranov equation (\cite{GR1958}, \cite{1966RvPP....2..103S}):
%%%%
\begin{equation}\label{gf}
a_1{''}e^{-2\Lambda}+\left(\Phi'-\Lambda' \right)e^{-2\Lambda}{a_1}'-\frac{2}{r^2}a_1=-4\pi j_1 \, \, ,
\end{equation}
%%%%
with the appropriate boundary and initial conditions, see \cite{2008MNRAS.385.2080C} for details.
As discussed in \cite{2009MNRAS.396.1441C},  $a_1(r)$ shows a maximum inside the star and then the transformation (\ref{new_trasf}) is not any more one to one. Therefore, after this maximum we choose the  minus sign in equation (\ref{new_trasf}).

In this paper we  extend our previous work incorporating a solid crust in the magnetar model. The presence of the crust is accompanied with a non-zero  shear modulus $\mu$ in the coefficients of the wave equation (\ref{2d}).    
The presence of the  shear modulus $\mu$  does not allow for dimensional reduction and the transformation of equation (\ref{2d}) leads again into  $2D$- wave equation, in the new  coordinates $X$ and $Y$. The new equation has the following form
%%%%%%%%
\begin{equation}\label{2d_crust}
\begin{split}
 A_{tt}\frac{\partial ^2 {\cal Y}}{\partial t^2}=&\bar{A}_{20}\frac{\partial^2 {\cal Y}}{\partial X^2}+\bar{A}_{11}\frac{\partial^2 {\cal Y}}{\partial Y \partial X}+\bar{A}_{02}\frac{\partial^2 {\cal Y}}{\partial Y^2}\\&+ \bar{A}_{01} \frac{\partial {\cal Y}}{\partial Y}+\bar{A}_{10} \frac{\partial {\cal Y}}{\partial X} ,
 \end{split}
 \end{equation}
%%%%%%
where the coefficients are 
%%%%%%
\begin{eqnarray}\label{coeff_2d}
\bar{A}_{02}&=&\frac{1}{4 \pi r^4}a_1 {a_1}'^{2}+\mu \pi r^2\biggl(\frac{{a_1}'^{2}}{4a_1}\cos^2 \theta\\
&+&a_1 e^{\Lambda}\sin^2 \theta \biggr),\\
\bar{A}_{20}&=&\mu \pi r^4 \biggl(\frac{{a_1}'^2}{4a_1}\sin^2 \theta+\frac{a_1 e^{\Lambda}\cos^2 \theta}{r^2}\biggr),\\
\bar{A}_{11}&=&\mu \pi r^4\biggl(\frac{{a_1}'^2}{2a_1}-2\frac{a_1 e^{\Lambda}}{r^2}\biggr)\sin \theta\cos \theta,\\
 \bar{A}_{01}&=&\frac{X}{2\pi r^4}a_1\left(\frac{2}{r^2}a_1 e^{2\Lambda}-4\pi j_1e^{2\Lambda}-\frac{{a_1}'^{2}}{2}\right),\\
 &+&{\mu}'\frac{{a_1}'^2}{2\sqrt{a_1}}\\&+&\frac{\mu \pi r^4}{\sqrt{a_1}}\biggl(\frac{{a_1}'^2}{a_1}-2\frac{a_1}{r}-\frac{a_1 e^{\Lambda}}{r^2}-2\pi J_1\biggr)\cos \theta , \\
 \bar{A}_{10}&=&\biggl[\frac{\mu }{\sqrt{a_1}}\biggl(-\frac{{a_1}'^2}{4a_1}+2\frac{{a_1}'}{r}-2\pi J_1\biggr)\sin \theta \\
                       &-&3\mu \sqrt{a_1}e^{\Lambda}\frac{\cos \theta}{r^2  }+{\mu}'\frac{{a_1}'^2}{2\sqrt{a_1}}\cos \theta \biggr]\pi r^4 \, .
\end{eqnarray}
%%%%%%
Note that for $\mu=0$ one recovers equation (\ref{1d_n}).
%The coefficients of the crust and the ones in the core  are all function of (X,Y)  as well as the shear modulus $\mu$. 
The value of $\mu$ in the following is calculated from  $v_s$, the shear velocity, see \cite{1983MNRAS.203..457S}
\begin{equation}
\label{mu}
v_s=(\mu/\rho)^{1/2} \simeq 1\times 10^8\;\; cm/s.
\end{equation}

%%%%%%%%%%%%%%%%%%%%
\subsection{Boundary Conditions}
%%%%%%%%%%%%%%%%%%%%

In order to solve the equations (\ref{1d_n}) and (\ref{2d_crust}),  appropriate boundary conditions must be imposed. 
Actually, the boundary conditions in spherical coordinates  were reported in  \cite{2007MNRAS.375..261S}. 

Since the wave equation is given in  the $(X,Y)$ coordinates  the aforementioned boundary conditions  translate as follows:
\begin{itemize}
\item regularity at the center: ${\cal Y}(X,Y)=0$; %for $X=0$ and $Y=0$,
\item no traction on the surface for the open lines:  
\begin{equation}\label{coupling}
\left[X\frac{\partial {\cal Y}}{\partial X}+ Y\frac{\partial {\cal Y}}{\partial Y}\right]=0; 
\end{equation}
\item axisymmetry % at $X=0$:
\begin{equation}\label{coupling1}
Y\frac{\partial {\cal Y}}{\partial X}=0  \quad \mbox{ at} \quad X=0; 
\end{equation}
\item equatorial plane symmetry for $\ell=2$ initial data i.e. %at $Y=0$:
%%%%
\begin{equation}\label{coupling2}
Y\frac{\partial {\cal Y}}{\partial X}=0  \quad \mbox{at} \quad Y=0; 
\end{equation}
%%%%%
\item equatorial plane symmetry for $\ell=3$ initial data i.e.  
%%%%%
\begin{equation}\label{coupling3}
{\cal Y}(X,Y)=0\quad \mbox{at} \quad Y=0;
\end{equation}
%%%%
\item continuity of the radial function $a_1$ when  the sign is switched in the coordinate transformation  (\ref{new_trasf}) at $a_1'(r)=0$.
\end{itemize}

An additional  boundary condition is needed at the base of the fluid-core interface: this condition in spherical coordinates is given by
\begin{equation}
\label{interface}
\partial_r{\cal Y}_{(-)}=\biggl[1+\frac{(2\ell-1)(2\ell+3)}{3(\ell^2+\ell-1)}\frac{v^2_s}{v^2_A}\biggr] \partial_r{ \cal Y}_{(+)}
\end{equation}
where $v_A=B/(4\pi \rho)^{1/2}$ is the Alfv\'en velocity, see \cite{2007MNRAS.375..261S} and \cite{2006MNRAS.371L..74G}. 
In the $(X,Y)$ coordinates the previous conditon becomes:
 \begin{equation}
 \begin{split}
\label{interface1}
&\left[X\frac{\partial {\cal Y}}{\partial X}+Y\frac{\partial {\cal Y}}{\partial  Y} \right] _{(-)}=\\ &\biggl[1+\frac{(2\ell-1)(2\ell+3)}{3(\ell^2+\ell-1)}\frac{v^2_s}{v^2_A}\biggr]\left[X\frac{\partial {\cal Y}}{\partial X}+ Y\frac{\partial {\cal Y}}{\partial Y}\right]_{(+)}.
\end{split}
\end{equation}
Although in this two dimensional study we have not used decomposition in spherical harmonics, we kept the relation (\ref{interface1})  in order to be able, during the mode recycling, to infer in the best possible way the form of the eigenfunction that corresponds to a specific angular index $\ell$. For the initial  runs we have used  the more generic formula:
\begin{equation}\label{interface2}
 \partial_r{\cal Y}_{(-)}=\biggl[1+v^2_s/v^2_A\biggr] \partial_r{ \cal Y}_{(+)}
  \end{equation}
which has been also used by \cite{2011MNRAS.410L..37G} to extract the general picture of the spectrum. In general, neither the spectrum nor the eigenfunctions  depend critically on this choice. 
%%%%%%%%%
\subsection{Numerical method}
%%%%%%%%%%

We used two representative EOSs the APR  \citep{1998PhRvC..58.1804A} and the WFF (\cite{1988PhRvC..38.1010W}),  for various mass models and different values of the magnetic field. At the end the ones that seem to fit better the observational data are the ones with a mass of $1.4M_\odot$  while the value for the   magnetic field strength which is in accordance to independent estimations was $B_\mu=4\times 10^{15}$ Gauss on the pole.

We used a numerical equidistant grid $60 \times 60$ in the
$(X,Y)$ coordinates, setting $X_{\rm max}=\sqrt{a_{1\; \rm max}}\sin \theta$ 
and $Y_{\rm max}=\sqrt{a_{1\; \rm max}}\cos \theta$ and varying 
$\theta$ from $0$ to $\pi/2$.  The accuracy of the code was tested, performing a simulation with a $90\times90$ equidistant grid: the results show that the frequencies are not significantly influenced by a change in the number of grid points. The stability of the scheme was also checked: our simulation lasted 2 seconds without showing any signs of instability.
The base of crust ($X_{\rm crust}$, $Y_{\rm crust}$ ) is taken at $\rho=2.4\times 10^{14}$ $g/cm^3$, according to the crust model by  \cite{1973NuPhA.207..298N} (NV).  In this way the study is divided into two  evolution problems coupled via the the interface condition (\ref{interface1}).

  First, we evolve the 1-dimensional wave equation (\ref{1d_n}) till the base of the crust  ($X_{\rm crust}$, $Y_{\rm crust}$). Special care should be taken at the point where the coordinate transformation (\ref{new_trasf}) is changing sign, i.e. at $a_{1\; \rm max}$. In practice, one evolves along ``magnetic strings'' throughout the core of the star till the base of the crust.
  Then in the crust, we evolve the 2-dimensonal wave equation (\ref{2d_crust}), and the oscillating ``magnetic strings'' of the core are replaced by a ``magneto-elastic membrane''.

As a test run for our model, we reproduce the results in \cite{2007MNRAS.375..261S} for crust oscillations. In this paper, the authors study the torsional oscillations of a non-magnetized star. They also consider, as a first approximation, the no-traction condition in the core-crust  interface, i.e. instead of the condition (\ref{interface}) they require that ${\cal Y}$ has to be continuous through the interface i.e.
${\cal Y}'_{(+)}={\cal Y}'_{(-)}$.
Here, we cannot set the magnetic field equal to  zero since our coordinate system $(X,Y)$ is based on the  magnetic field strength i.e. on $a_1(r)$. 
However, it was shown in \cite{2007MNRAS.375..261S} that the influence of the magnetic field on the frequencies becomes important only if $B>10^{15}$ Gauss, thus 
we can choose a very low magnetic field for magnetars, e.g. $B=10^{14}$ Gauss, and perform test runs in order to reproduce the results of \cite{2007MNRAS.375..261S}. Using this magnetic field and the no-traction condition on the interface, we find the results reported in \cite{2007MNRAS.375..261S}. 
This test verified that the code can reproduce earlier results for the torsional oscillations of the crust.  

\begin{figure}
\begin{center}
%\epsfxsize = 7in
% \mbox{\psfig{evolution.ps,width=8.0cm,angle=-90}}

\includegraphics[width=0.47\textwidth]{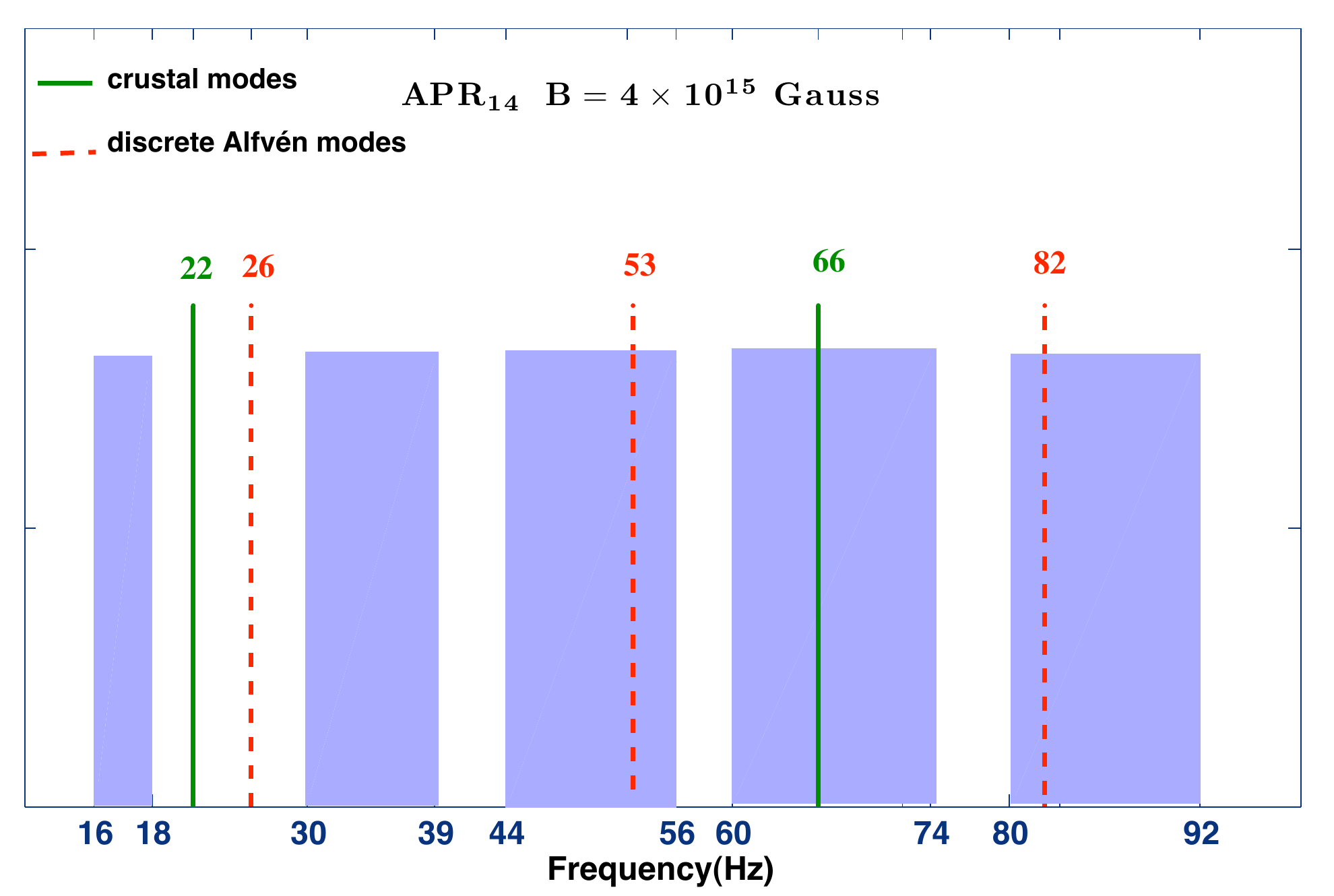}
\includegraphics[width=0.47\textwidth]{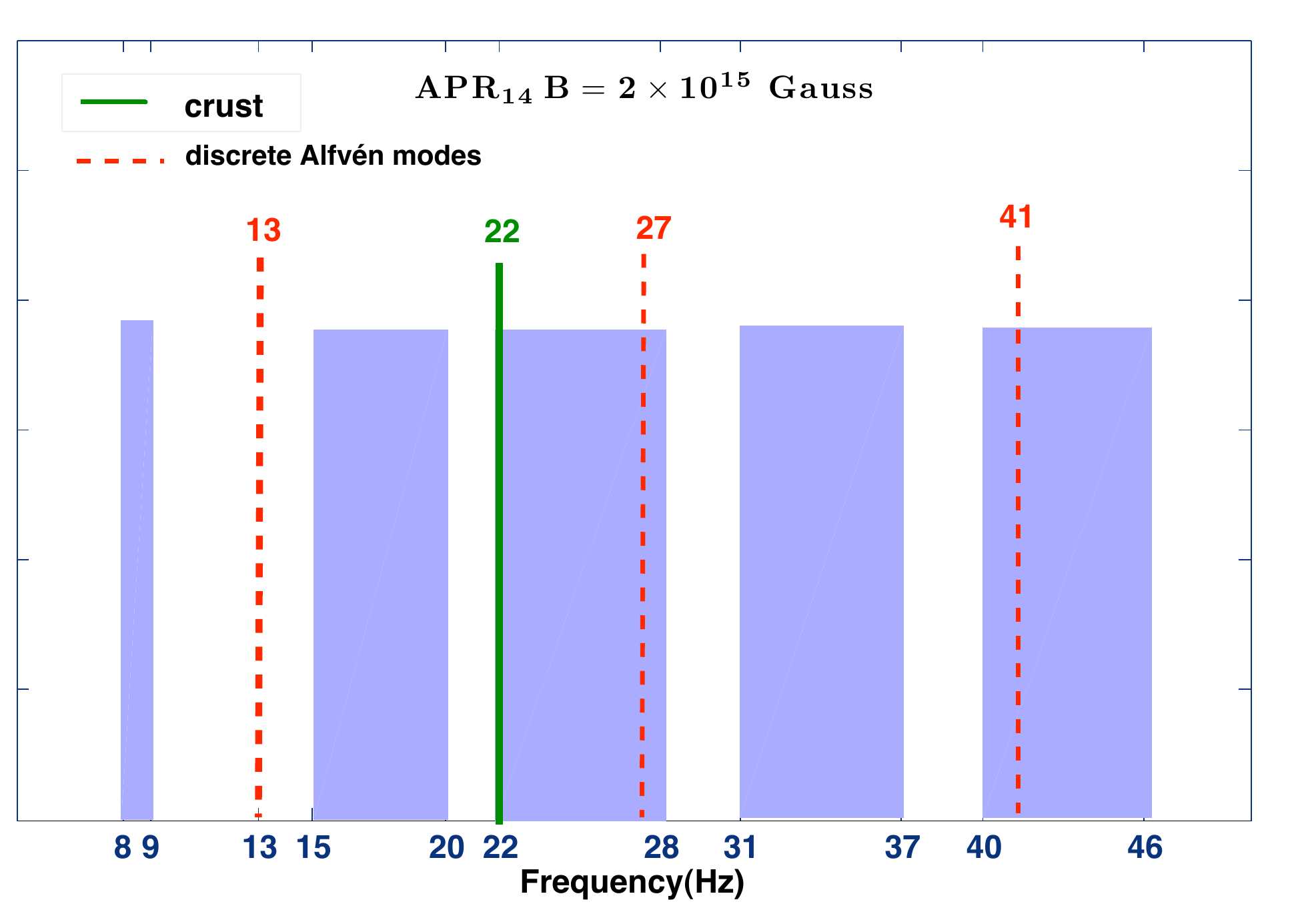}
\caption{%%
Identification of the frequencies of SGR 1806-20. We show that a stellar model APR  with mass $M=1.4 M_{\odot}$ and radius $R=11.57$km can explain all the frequencies observed. {\em Upper panel:} the  magnetic field strength is $4\times 10^{15} G$. {\em  Lower panel:} the  magnetic field strength is $2\times 10^{15} G$.
 In both panels the blue bands represent the continuous spectrum of the core oscillations.}%
\label{FFT_APR}
\end{center}
\end{figure}

\begin{figure}

\begin{center}
%\epsfxsize = 7in
% \mbox{\psfig{evolution.ps,width=8.0cm,angle=-90}}

\includegraphics[width=0.47\textwidth]{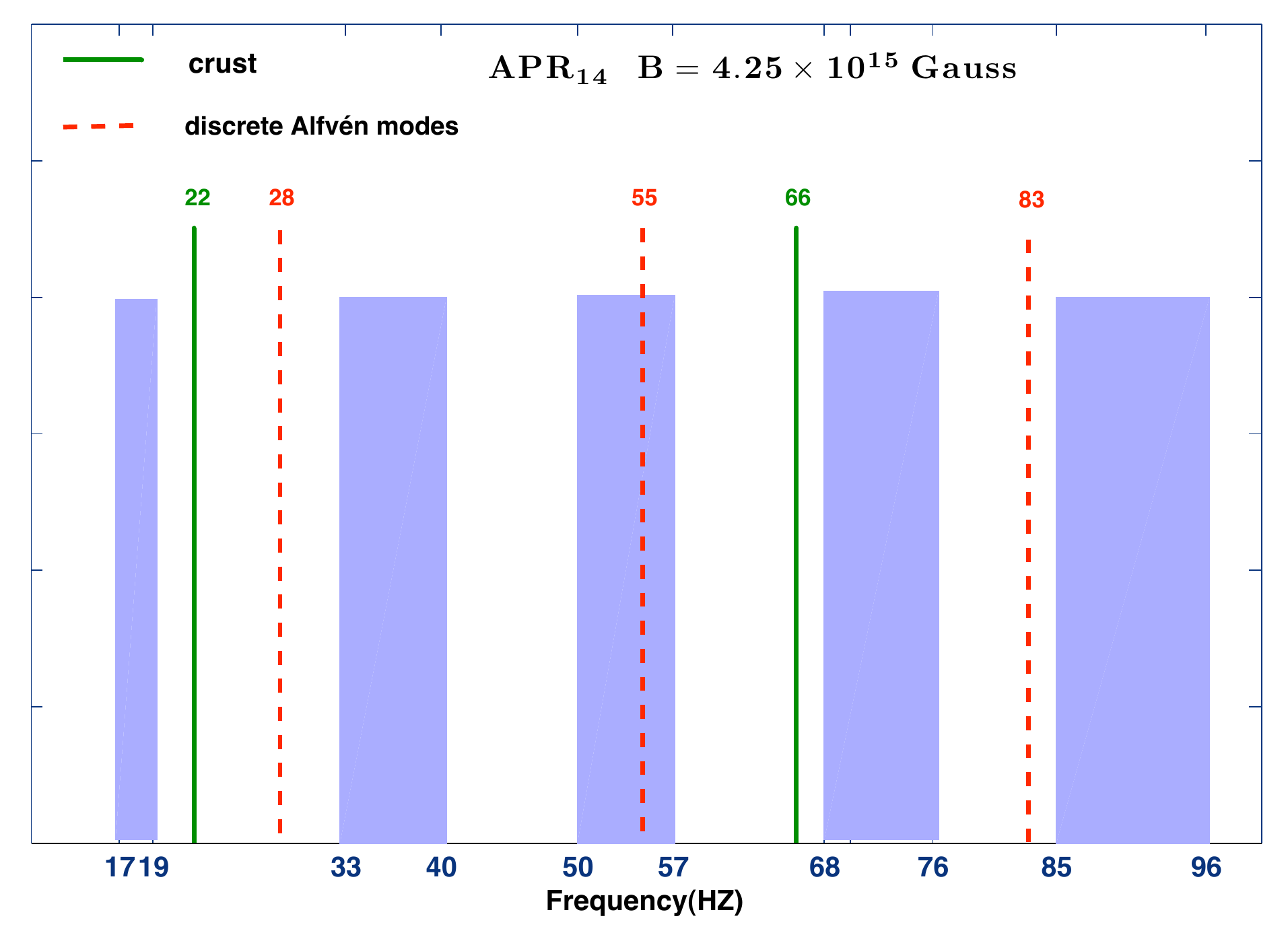}
\includegraphics[width=0.47\textwidth]{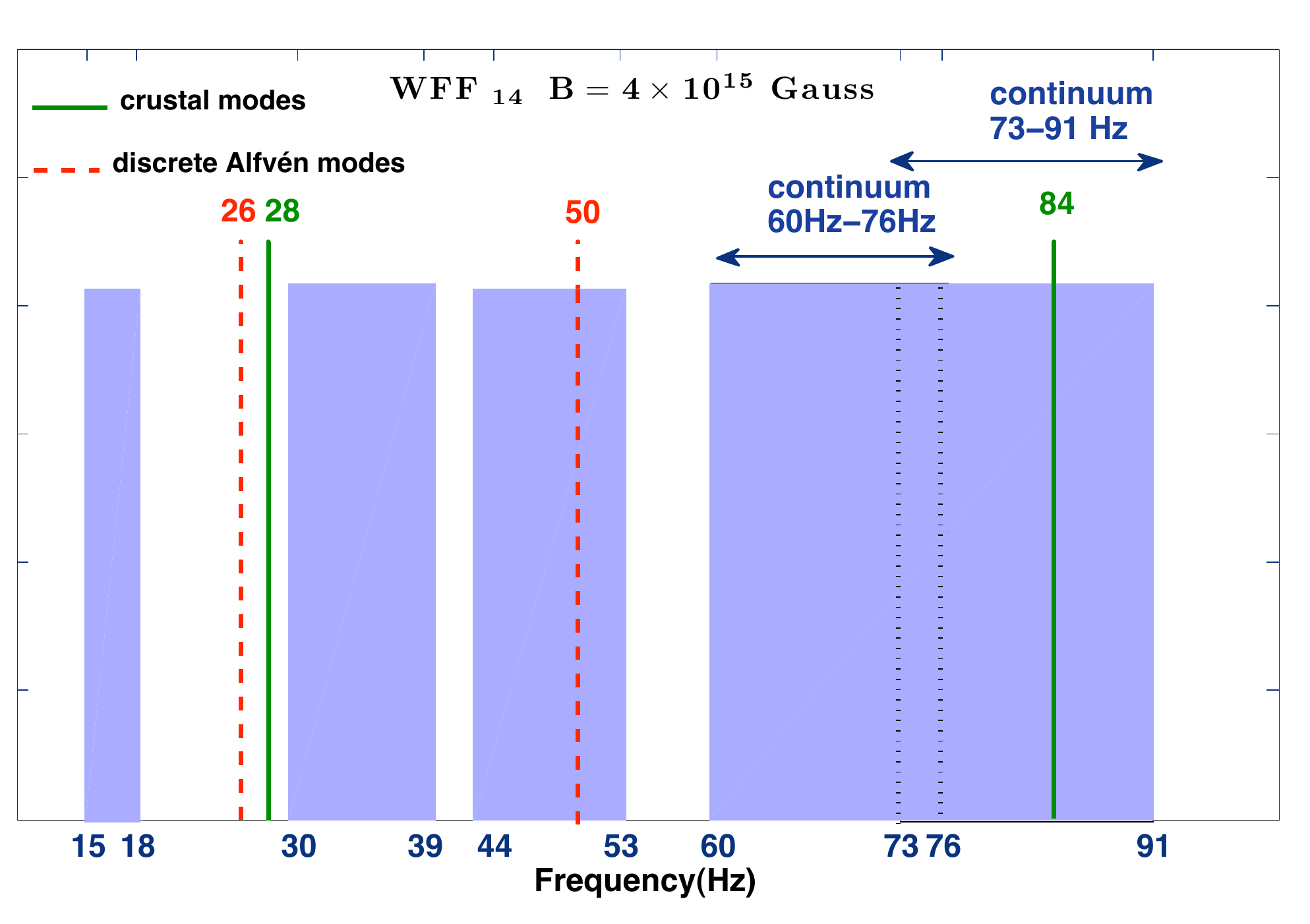}
\caption{%%
Identification of the frequencies of SGR 1900+14. Since just  few frequencies were identified for this SGR,  we show that a stellar model APR ({\em upper panel}) with mass $M=1.4 M_{\odot}$, radius $R=11.57$km  and magnetic field $B=4.25\times 10^{15}$G can explain all the frequencies observed, as well as  a stellar model WFF with mass $M=1.4 M_{\odot}$, radius $R=10.91$km  and magnetic field $B=4\times 10^{15}$G ({\em lower panel}).  In both panels the blue bands represent the continuous spectrum of the core oscillations.  
}%
\label{FFT_WFF}
\end{center}
\end{figure}

In a similar way we can isolate the continua  found in \cite{2009MNRAS.396.1441C}  and trace its changes due to the presence of the crust. One of the characteristic properties of the continua is the scaling of the  frequencies that seem to form the edges of the continua. That is,  the edges of the various continua follow the rule that has been found for odd type of initial data  in \cite{2009MNRAS.396.1441C} i.e. that 
%%%
\begin{equation}
\label{continuum_scale}
f_{L_n}\approx (n+1) f_{L_0} \quad \mbox{and} \quad f_{U_n}\approx (n+1) f_{U_0},
\end{equation}
 %%%
where $f_L$ stands for the lower frequency of the continua with an eigenfunction located mainly near the polar axis, and   $f_U$ stands for the upper frequency of the continua with maximum amplitude near the last open magnetic field line. These scaling laws for the continua can be easily spotted in Figs.   \ref{FFT_APR} and \ref{FFT_WFF}. 
 
In this coupled core-crust problem  apart from the continua and the crust modes, we found the imprints of a new family of   discrete modes, a detailed description of these modes  is presented in the following section. 

 %-----------------------------------------------------------------------------
%---------------------------------------
%-----------------------------------------------------------------------------
%%%%%%

\begin{figure}
\begin{center}
%INTERFACE
\includegraphics[width=8cm, height=7cm]{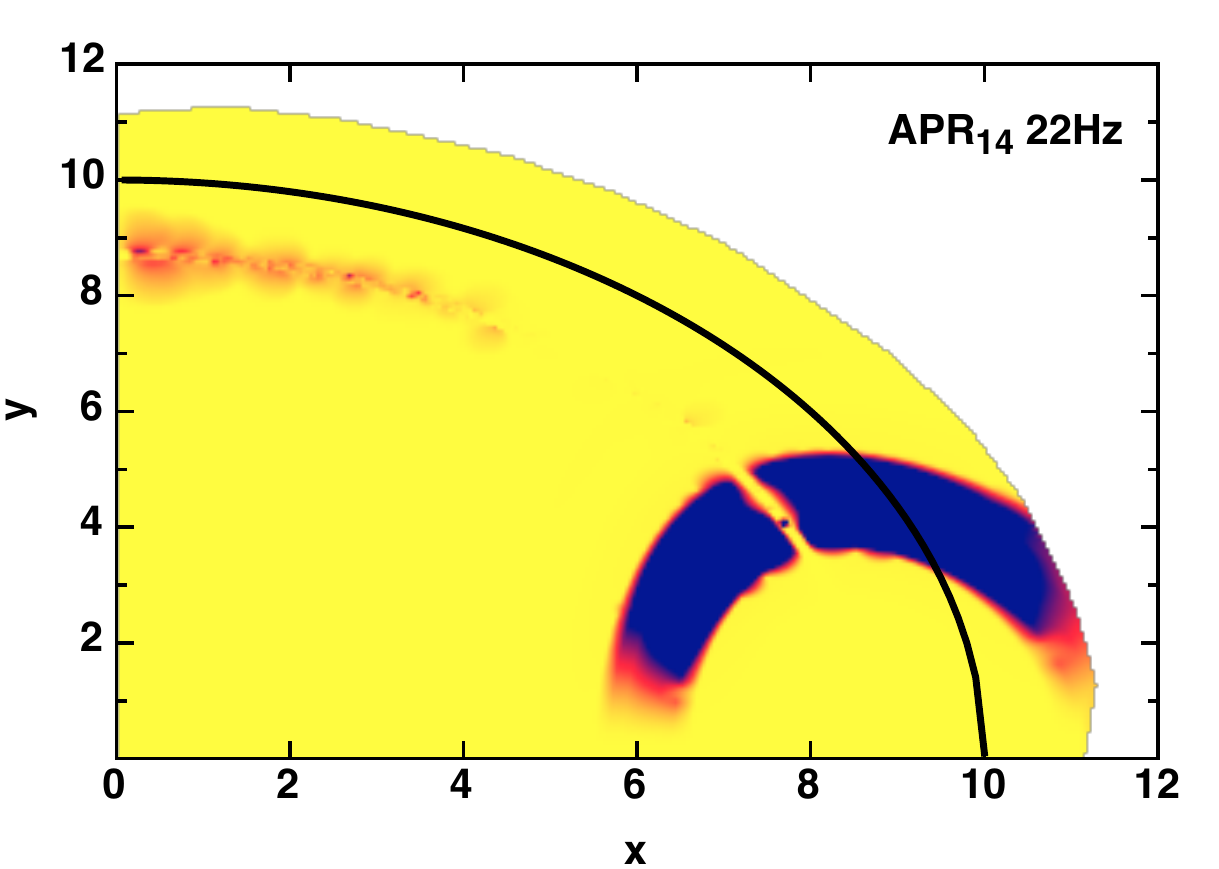}
\includegraphics[width=8cm, height=7cm]{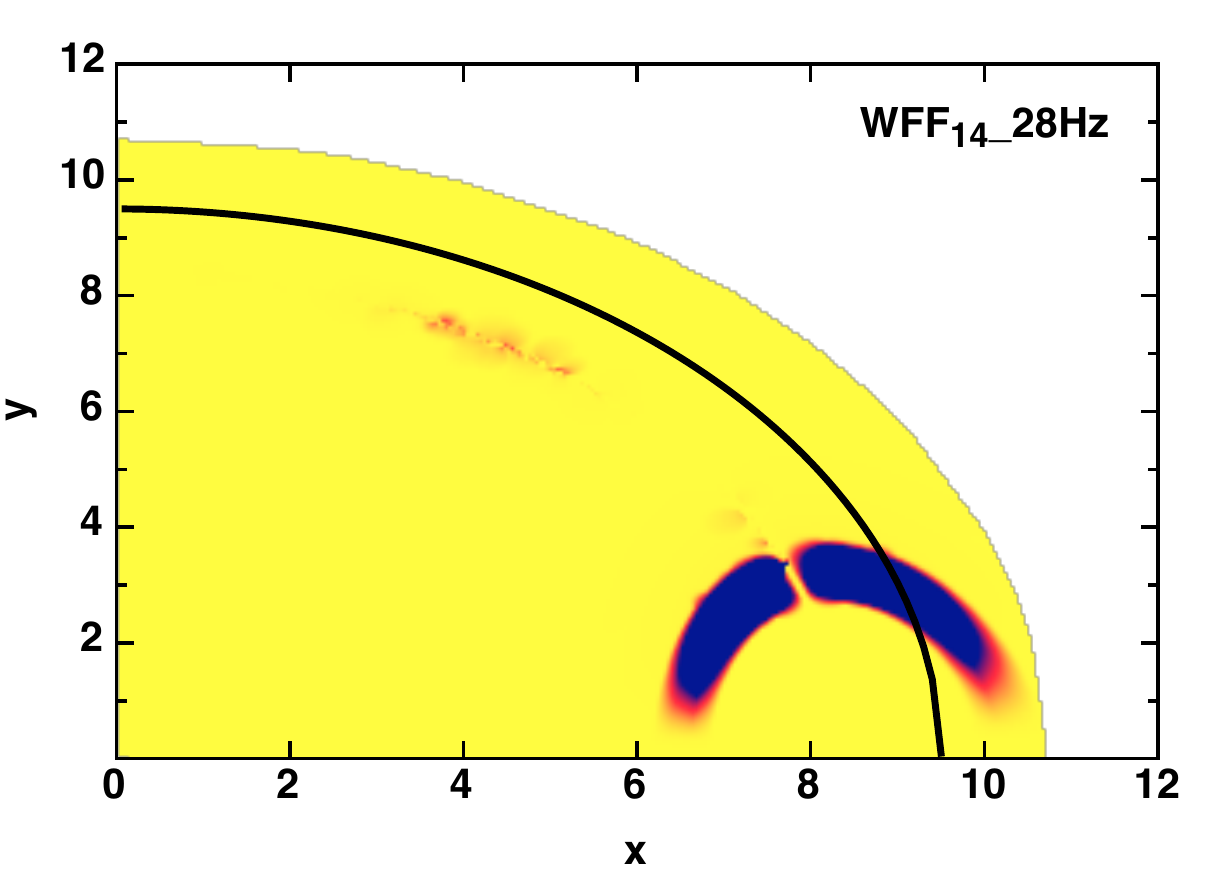}
\caption{%%
Typical eigenfunctions of a ``crust'' mode  for the model APR$_{14}$ that identifies the SGR 1806-20 ({\em upper panel}) and the model WFF$_{14}$ ({\em lower panel}), that identifies the SGR 1900+14. This latter frequency is unlike to be observed, since it does not reach the stellar surface.  The noise visible in  both figures corresponds to the change of sign of the $(X,Y)$ coordinate at the point  $a_{1\rm max}$.
}%
\label{crust_APR}
\end{center}
\end{figure}

\begin{figure}
\begin{center}
%CRUST
\includegraphics[width=8cm, height=7cm]{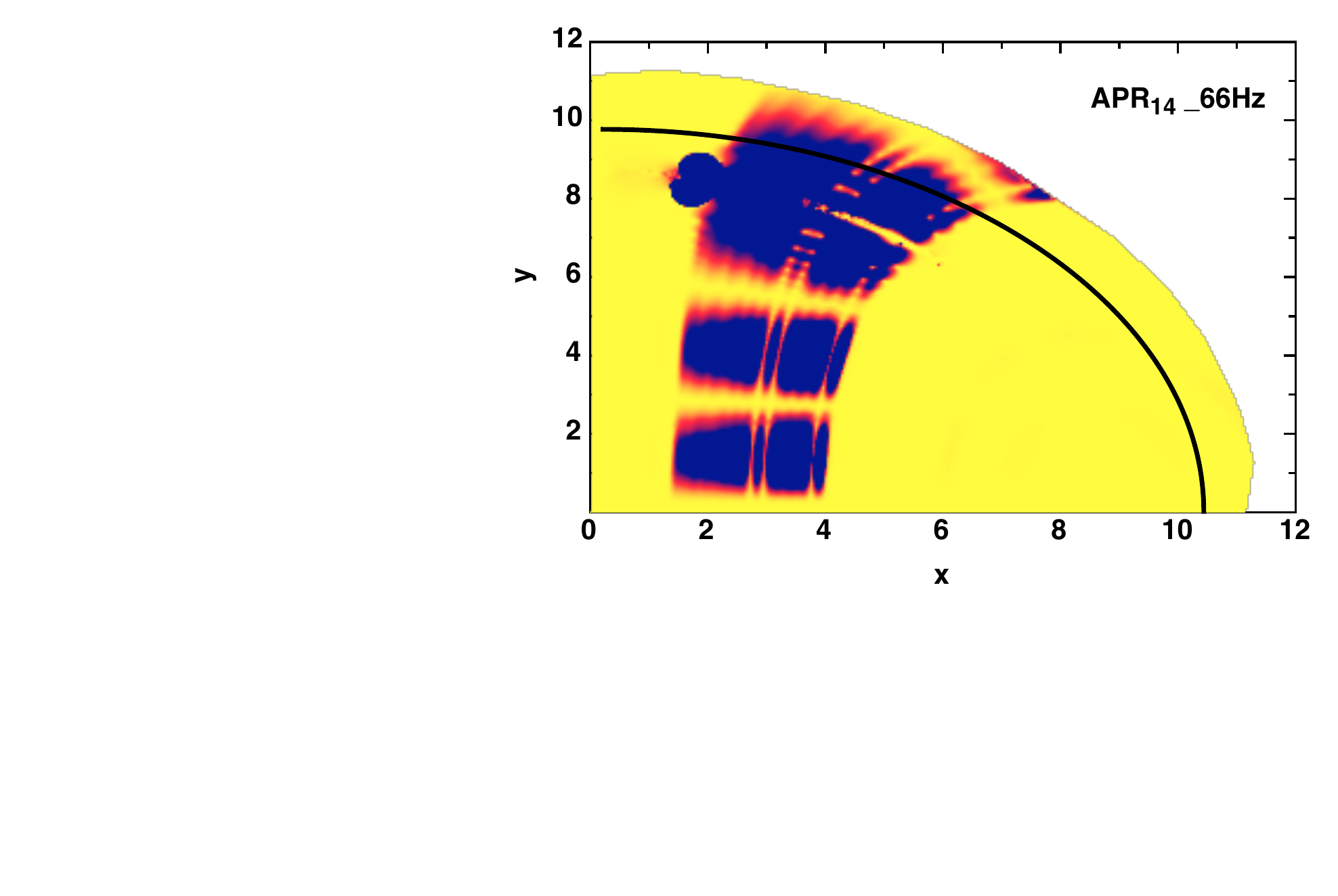}
\includegraphics[width=8cm, height=7 cm]{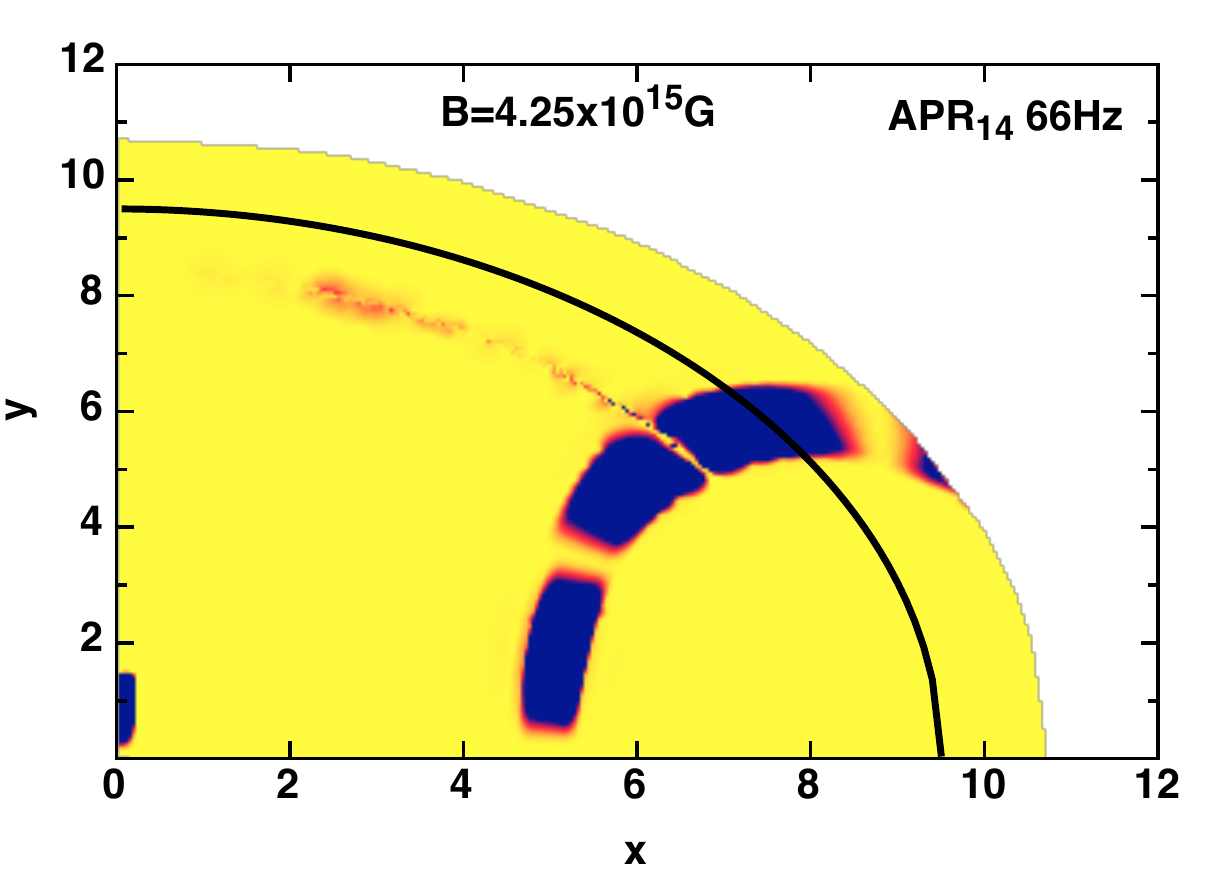}
\caption{%%
Eigenfunctions of the ``crust'' mode 66Hz for the APR$_{14}$. {\em Upper panel:} the  the magnetic field strength is $B=4\times 10^{15} G$  and the frequency is hosted in the continua (model for SGR 1806-20). {\em Lower panel:} the frequency is  located in a gap and the magnetic field is $B=4.25\times 10^{15} G$ (model for SGR 1900+14).
 }%
\label{eigen_66Hz_APR}
\end{center}
\end{figure}

Finally, we made some additional numerical tests i.e we set initial data in order to  excite only  the crust and to study the propagation of the energy in the system. 
We pick up two points, one in the core and one in the crust,  and we follow the evolution of the perturbation during the time of our simulation, performing a FFT at different times and comparing the amplitude of their peaks.
 We found that the energy quickly flows from the crust towards the core, exciting the Alfv\'en continua as seen by \cite{2007MNRAS.377..159L} and by \cite{2011MNRAS.410L..37G}. In the same way, the perturbations of the core excite the crust: for this reason it is more appropriate to define the modes generated by the continua, global modes, i.e. modes that involve both  crust and core oscillations.
The conclusion of this test is in agreement with earlier suggestions \citep{2006MNRAS.371L..74G} claiming that 
when the magnetic field permeates the whole the star,  the perturbations cannot be confined in the crust.
Another notable result is that even when we  associated a mode to the crust, the actual eigenfunction was not confined in the crust but it was extended throughout the star, see Fig.  \ref{crust_APR}.  
In addition, we find that when a crustal mode is embedded in the continua, its eigenfunction shows the same structure of the continua that hosts it. Contrary, when a crust frequency is found in the gap between the continua, its eigenfunction shows a structure more similar to the one of the fundamental crustal frequencies, compare   the upper panel of Fig. \ref{crust_APR} and the lower panel of Fig.  \ref{eigen_66Hz_APR}.

 %-----------------------------------------------------------------------------
%---------------------------------------

%-----------------------------------------------------------------------------
%---------------------------------------
   %-----------------------------------------------------------------------------
\begin{figure}
\begin{center}
%INTERFACE
\includegraphics[width=0.47\textwidth]{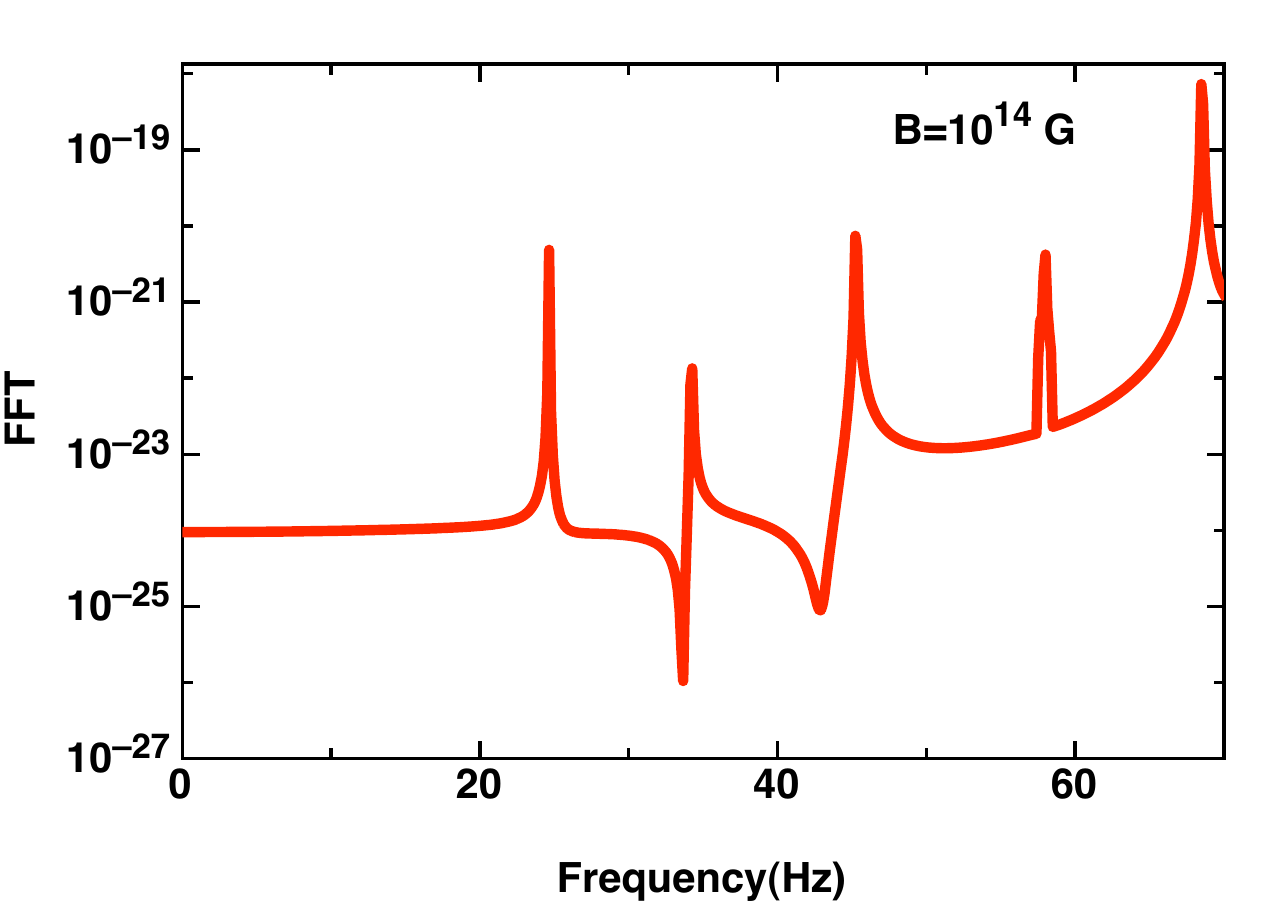}
\includegraphics[width=0.47\textwidth]{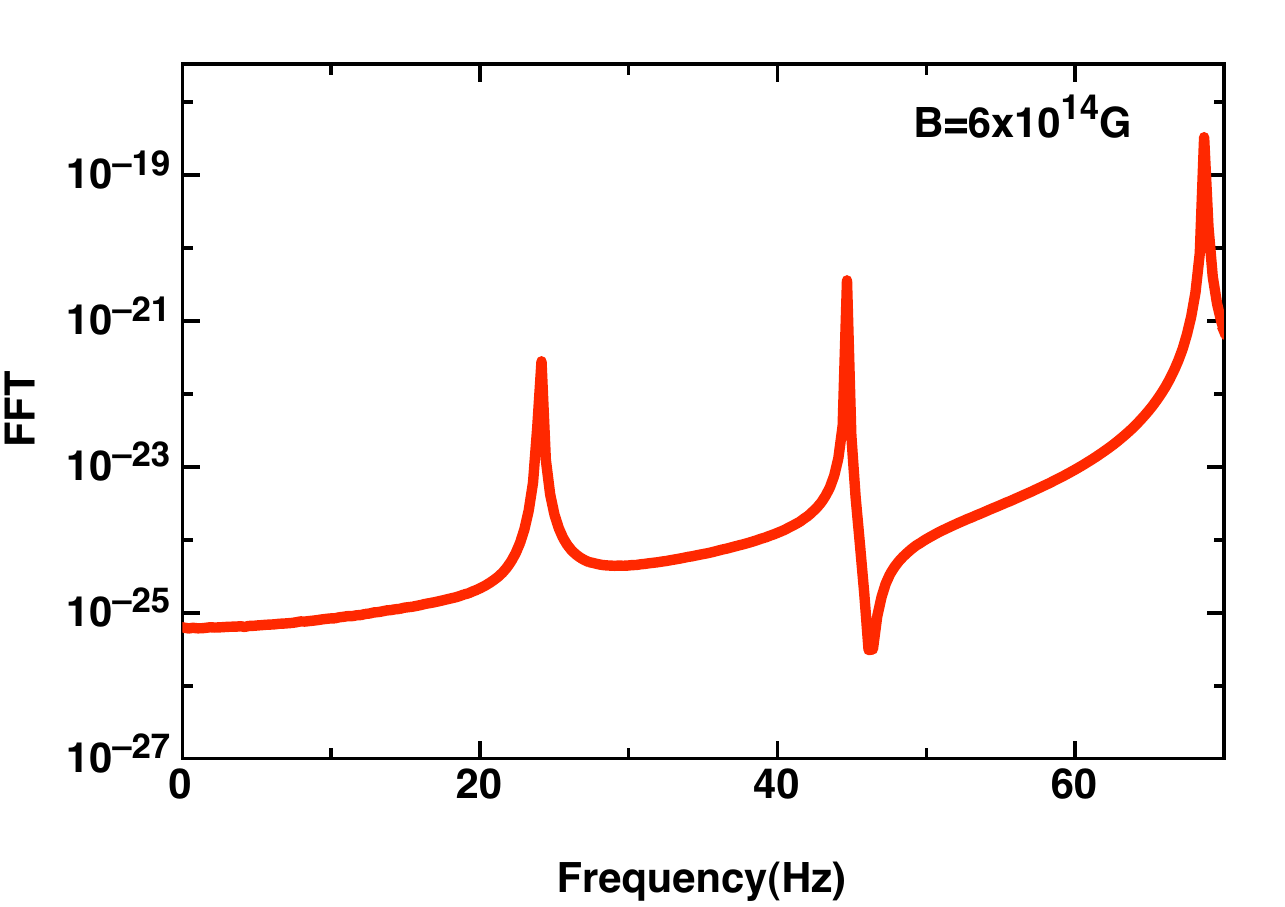}
\caption{%%
{\em Upper panel}: the crustal frequencies of an APR$_{14}$ model for a weak magnetic field $B\simeq 10^{14}$ Gauss. {\em Lower panel}:  for a weak magnetic field $B\simeq 6\times 10^{14}$ Gauss: some of the frequencies disappear as the magnetic field increases.
 }%
\label{lowB}
\end{center}
\end{figure}
%---------------------------------------

%%%%%%%%%%%%%%%%%%%%%%%%%%%%%%%%%%%%%%%%%
%-----------------------------------------------------------------------------
\begin{figure}
\begin{center}
%CRUST
\includegraphics[width=8cm, height=7cm]{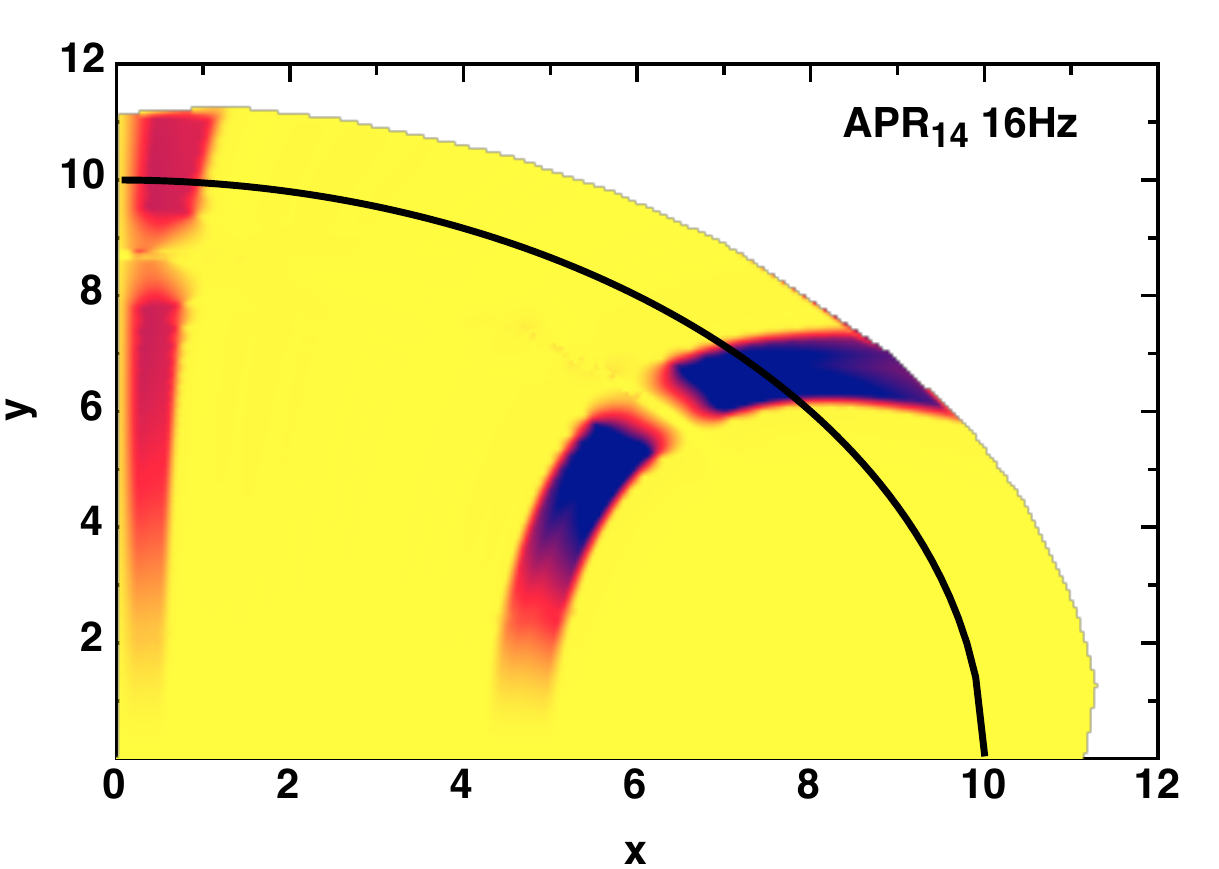}
\includegraphics[width=8cm, height=7cm]{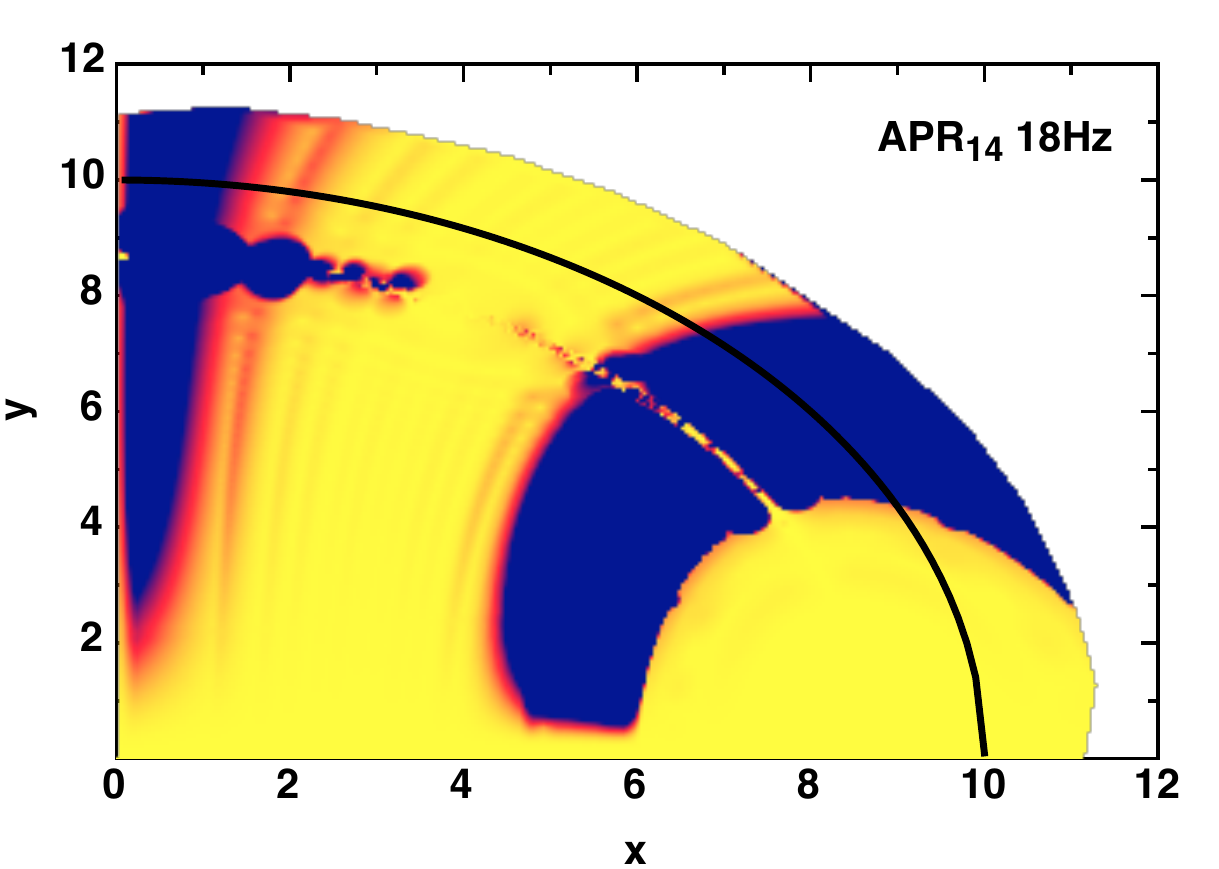}
\caption{%%
Eigenfunctions of the first two edges of the first continuum for the model  APR$_{14}$ (SGR 1806-20).
 }%
\label{eigen_c16-18_APR}
\end{center}
\end{figure}
%---------------------------------------

 %-----------------------------------------------------------------------------
\begin{figure}
\begin{center}
%INTERFACE
\includegraphics[width=8cm, height=7cm]{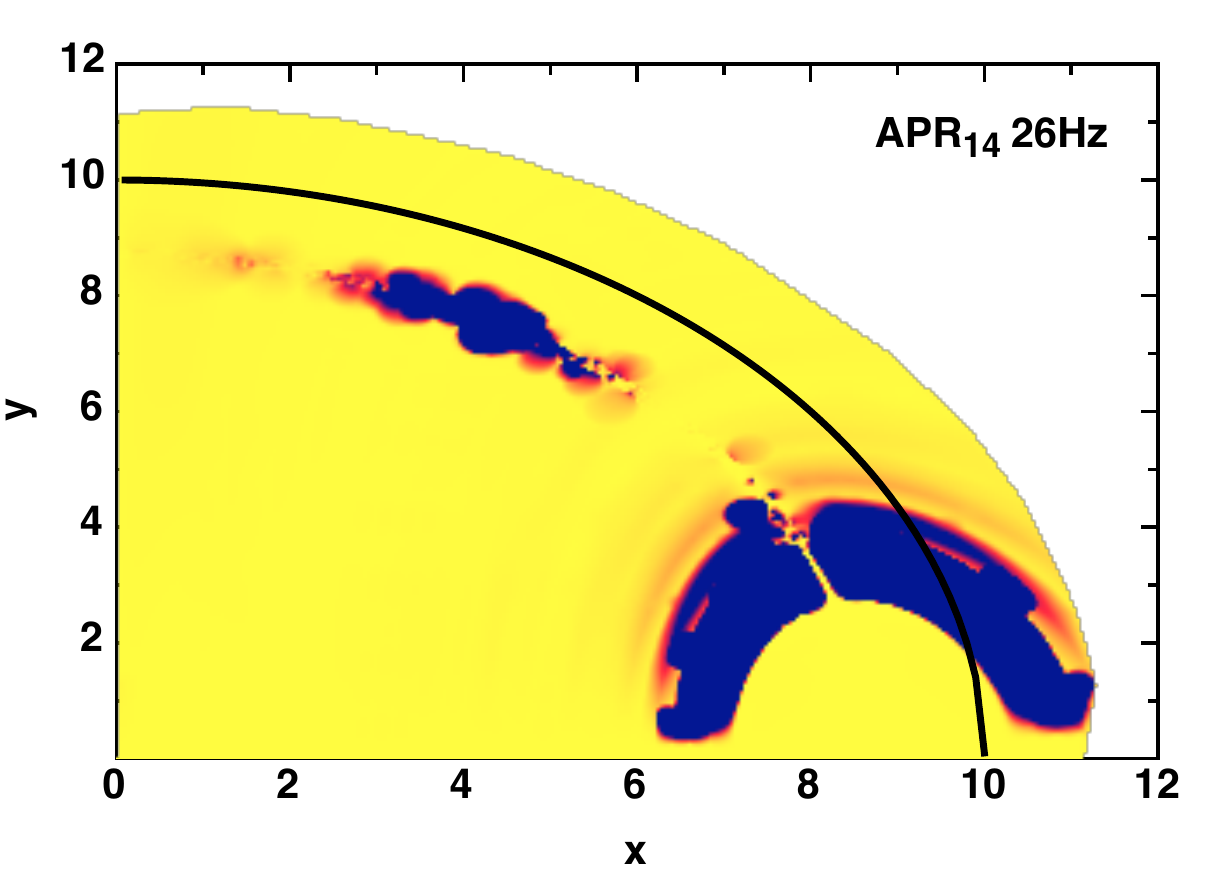}
\includegraphics[width=8cm, height=7cm]{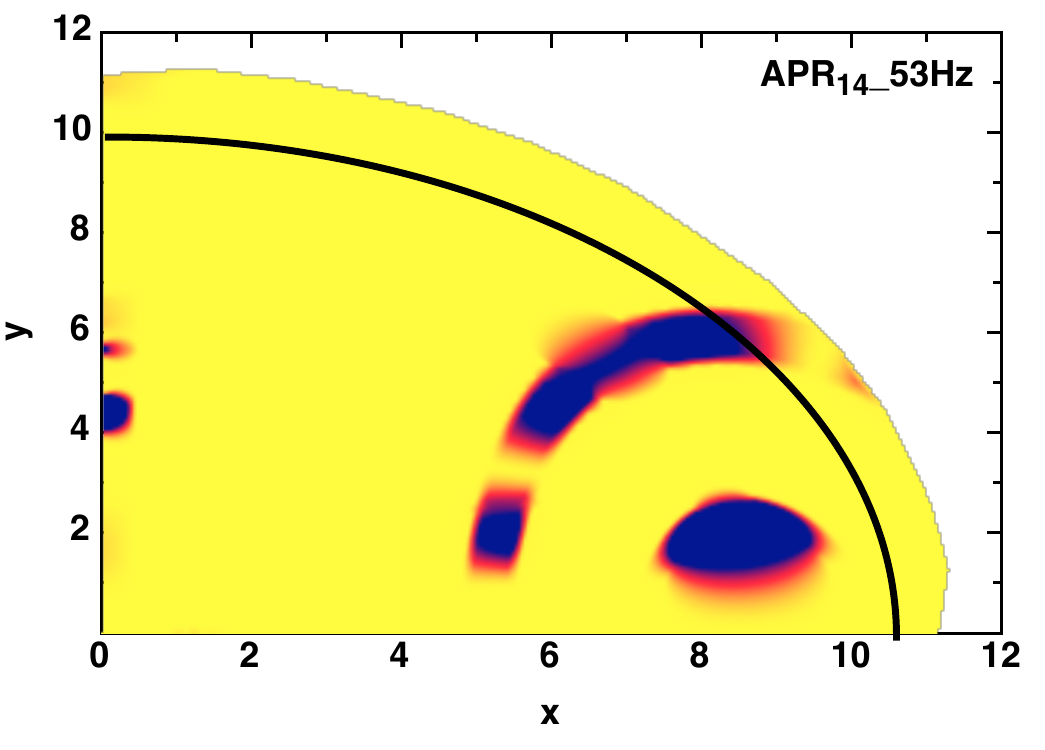}
\caption{%%
Typical eigenfunctions of a ``discrete Alfv\'en mode ''  for the model  APR$_{14}$ (SGR 1806-20). Note that the 53Hz frequency, represented in the lower panel, is unlike to be observed, since it does not reach the stellar surface.
 }%
\label{discrete_APR}
\end{center}
\end{figure}
%---------------------------------------

%-----------------------------------------------------------------------------
\begin{figure}
\begin{center}
%Continuum
\includegraphics[width=8cm, height=7cm]{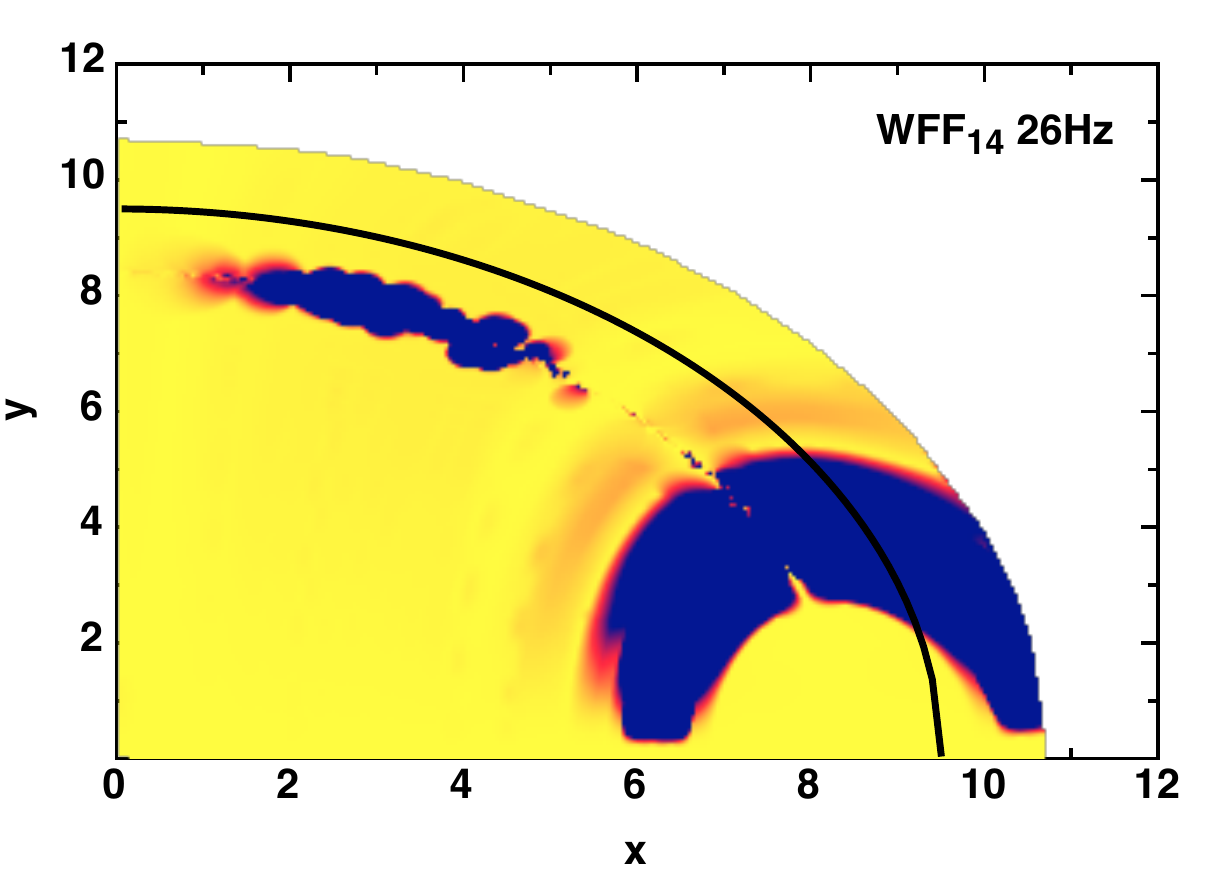}
\includegraphics[width=8cm, height=7cm]{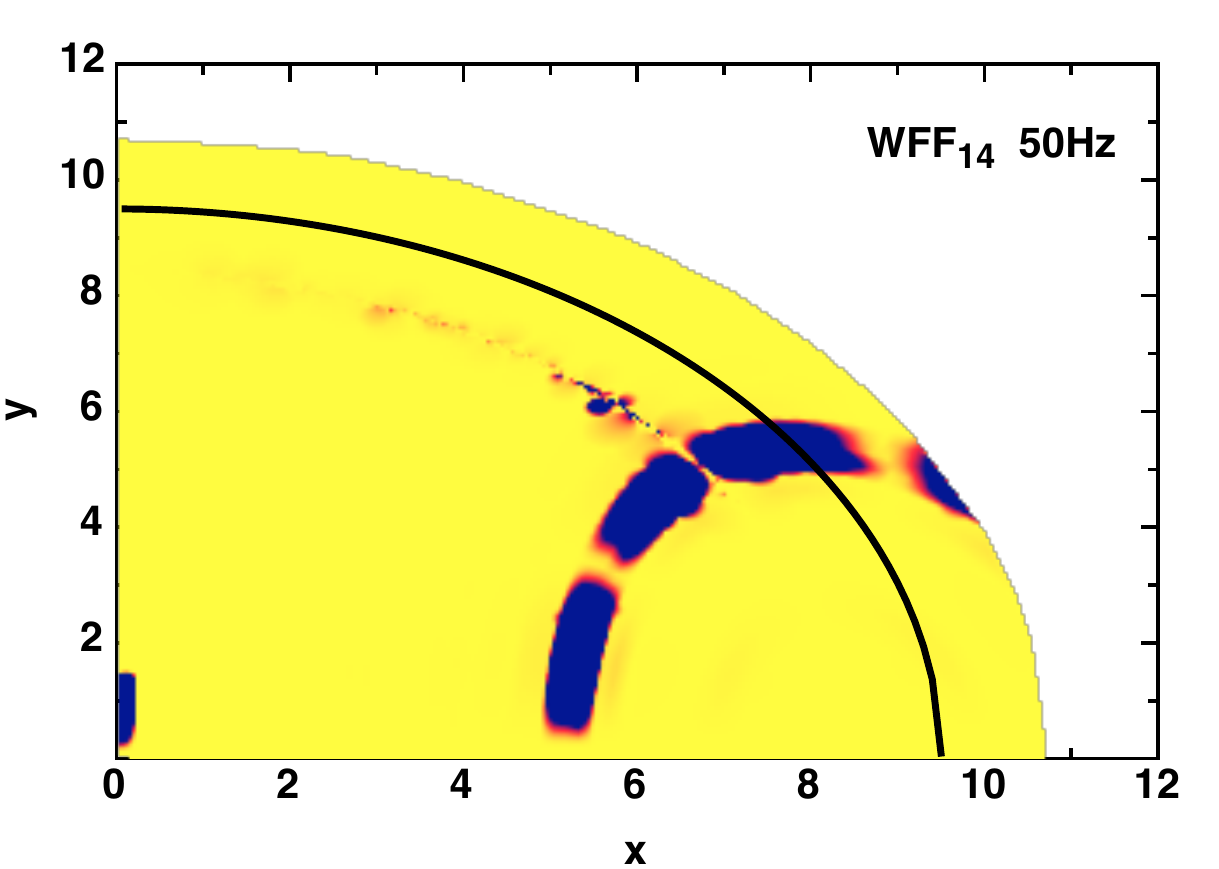}
\caption{%%
Typical eigenfunctions of a ``discrete Alfv\'en mode ''  for the model WFF$_{14}$ (SGR 1900+14).
 }%
\label{discrete_WFF}
\end{center}
\end{figure}
%---------------------------------------

%---------------------------------------

%%%%%%%%%%%%%%%%%
\section{Results \& Discussion}
%%%%%%%%%%%%%%%%%
The scope of this paper is not only to understand if the oscillations are global modes (i.e. combined crust-core oscillations), pure crustal modes or pure core modes  but also  to identify the magnetar models that can fit the observed QPOs.
This can potentially lead to a better understanding of the order with which the various modes were excited and left their imprints in the signal. Moreover, a proper identification of the observed QPOs will unveil the details of the magnetar structure. 

%%%%%%%%
\subsection{Analysis of the spectrum}
%%%%%%%%

As we already pointed out and as it is shown in Fig.  \ref{FFT_APR} and  Fig.  \ref{FFT_WFF}, the spectrum of the oscillations is composed by three different types of modes: 
\begin{description}
  \item[ \bf crustal modes:] they are associated with the crust and they acquire a discrete spectrum. We found that, when the magnetic field increases, some  modes are absorbed by the continua, compare upper panel of Fig.  \ref{lowB},  where $B\simeq 10^{14}$ Gauss, with the lower panel of the same figure, where $B\simeq 6\times 10^{14}$ Gauss: it is clear that some modes disappear. The strong coupling with the crust leads to an effective transfer of energy  from the crust to the core, especially when the crust frequency is embedded in the continua (see for example the 66Hz frequency in the upper panel of  Fig.  \ref{FFT_APR}): from a combined  analysis of the peak amplitude of a crust frequency  embedded in the continua and of the peak amplitude of  the edge of the hosting continua, we found that the energy lost by the crust frequency is stored in the edges of the continua.  In addition, we found that the fundamental crustal frequency does not scale with the increase or decrease of the magnetic field: this behaviour assures us that the crustal frequencies are not scaling with the magnetic field unlike the Alfv\'en modes. For magnetic field $B>10^{15}$ Gauss, the relation among the fundamental frequency and the overtones, is given by:
  \begin{equation}
\label{crust_fre2}
f^{\rm crust}_{\ell}=(\ell-1) f^{\rm crust}_{\ell=2} \quad \mbox{for} \quad \ell>2 \quad \mbox{and} \quad  n=0.
\end{equation}
For a magnetic field $B<4\times 10^{14}$ Gauss, the law found by  \cite{SA2007}  is recovered: 
  \begin{equation}
\label{crust_fre}
f^{\rm crust}_\ell=\sqrt{(\ell-1)(\ell+2)} \frac{v_s}{2\pi R}\quad \mbox{for }\quad n=0.
\end{equation}  
For intermediate magnetic fields, a proper scaling cannot be found. The problem of the scaling of the crustal mode was treated also by \cite{2011MNRAS.410L..37G}, where the authors discuss the appearance or disappearance of the crustal mode in relation to the magnetic field strength.
  \item[\bf continuous spectrum:] the continuous spectrum is generated by the presence of a strong magnetic field ($B>10^{15}$ Gauss). It is generated in the magnetised fluid present in the core but, in the case of strong magnetic field it extends and, through the coupling with the crust, its oscillations penetrate also in the crust, giving rise to global modes. When the magnetic field increases, the continuous spectrum becomes more and more energetic, absorbing also some crust modes. In addition, the edges of the continua scale with the magnetic field, since $f=B/(\sqrt{4\pi \rho} L)$, where $L=L(X)$ is the length of the ``magnetic strings'' along which the perturbation propagates (compare the upper and lower panel of Fig.  \ref{FFT_APR}).  The frequencies of the continua show the scaling described by equation (\ref{continuum_scale}).
  Note that the continuous spectrum always  shows gaps in its structure for the model APR (see Fig.  \ref{FFT_APR}), while the model WFF does not have a gap between the fourth and fifth continua  that overlap  (see Fig.  \ref{FFT_WFF}) : the presence/absence of gaps seems then to be a characteristic of a particular model more than a general one, as pointed out in \cite{2011MNRAS.410.1036H}.
  
    \item[\bf discrete Alfv\'en modes: ] these modes, as the crustal modes, are discrete but we identify them as Alfv\'en modes, because they scale with the magnetic field as the  frequencies of the continua. Their structure is similar to  the one of the crustal mode (compare the upper panel of Fig.  \ref{discrete_APR} and Fig.  \ref{discrete_WFF} with Fig.  \ref{crust_APR}) but their scaling follows the scaling observed in the continua:
 %%%
\begin{equation}
f_{n}^{\rm (D)}\approx (n+1) {f_{0}}^{\rm (D)} \, .
\end{equation}
 %%%   
  Note that those modes have not been observed in the case of the absence of  a crust (see \cite{2009MNRAS.396.1441C}) . For this reason, they can be  interpreted as an effect of the coupling between the  fluid core and a solid crust  (see \cite{2011MNRAS.410.1036H}).
\end{description}
 %%%%%%%%%
\subsection{Identification of the QPOs}
%%%%%%%%

From the variety of magnetars models that we have examined in order to fit to the observational data only a few provide oscillation frequencies that are in agreement with the observed QPOs.
More specifically, we compared the data that our numerical code produced with the frequencies from the timing analysis of the SGR 1806-20 and  SGR 1900+14 shown  in  \cite{SW2006}. In this paper,  the authors identify several QPOs, of different duration, in the tail of the two events. 
In particular, for the more recent event, SGR 1806-20, the identified frequencies are 18, 26, 30, 92, 150, 625 and 1840 Hz  
while for the SGR 1900+14 they found the following frequencies:  28, 53, 84 and 155 Hz.  
A more recent study by  \cite{2011A&A...528A..45H}, based on predictions by \cite{2009MNRAS.396.1441C},   confirms, by using a different analysis technique,  the earlier results and, in addition, unveils the presence of at least three new frequencies for the SGR 1806-20 these are 16, 21 and 36 Hz.  

The new results pose extra challenge, since already the previous studies of pure torsional oscillations in magnetars could not explain all observed frequencies, in particular the lower ones (18, 26, 29 Hz), mainly because of the small spacing among them.  If the two new frequencies (16Hz and 21Hz) are added then it is impossible by any means the explanation via crust oscillations alone. On the other hand, the Alfv\'en continua in the core \citep{2009MNRAS.396.1441C,2009MNRAS.397.1607C}  offer better chances for an explanation but still both calculations did not take into account the presence of a crust. 

Here we show that our modelling of magnetar dynamics can explain all the observed frequencies, and inversely via this explanation we can constrain the parameters of the magnetar (radius, mass, equation of state and magnetic field strength).
In Figs.   \ref{FFT_APR} and \ref{FFT_WFF} we graphically show our findings, which comprised by 3 types of oscillations, the Alfv\'en continua in the core, the discrete Alfv\'en modes (whose eigenfunctions are plotted in Figs.  \ref{discrete_APR} and \ref{discrete_WFF})  and the crustal torsional modes confined mainly in the crust (whose eigenfunctions are plotted in Fig.  \ref{crust_APR}).
 The eigenfunctions were found  using the eigenfunctions recycling program already used in \cite{2009PhRvD..80f4026G} (see there for more details).

 A model with the equation of state (EOS) APR for the core and the NV model for the crust,  with mass $M=1.4  M_{\odot}$ and radius $R=11.57$km, could fit the new data discovered by \cite{2011A&A...528A..45H} as well as the previous data for the SGR 1806-20 \citep{2007AdSpR..40.1446W}.  
 Here, we focus on the first five frequencies, i.e 16, 18, 22, 26, 29 Hz, since they are the more difficult to be explained, the higher ones can be explained in various ways, as multiples of the lower ones or as  polar modes (see \cite{2009MNRAS.395.1163S}).

 In the upper panel of Fig.  \ref{eigen_c16-18_APR}, the eigenfunction of the 16Hz frequency, i.e. the first edge of the continua (see  Fig.  \ref{FFT_APR}), is shown. The perturbation involves both core and crust and it can be classified as a global oscillation, with the crust and the core oscillating together due to the strong coupling  induced by the magnetic field. The perturbation seems to be localised near the magnetic axis and just outside the region with the closed lines. 
 In the bottom panel of Fig.  \ref{eigen_c16-18_APR}, the eigenfunction of the second edge of the first continuum, the 18Hz frequency, is shown: its structure is similar to the one of the 16Hz frequency, but it shows broader spreading and intense oscillation.

The  $22$ Hz mode has a discrete nature, as we already pointed out in the previous section, see upper panel in Fig.  \ref{crust_APR}. Studying the time evolution of the perturbation, as explained in the previous section, we find that the core is excited after some time, when the crust successfully  transfers its energy to it. It seems to be a fossil of crust mode found for weaker magnetic field and it is then reasonable to classify it as crustal frequency.
The next frequency found, at $26$ Hz, see Fig.  \ref{discrete_APR}, is similar in  structure to the crustal one but, analysing its time evolution, we find that it  originates at the crust core interface and its nature is discrete. In addition, although it is a discrete frequency, it scales with the magnetic field: it has  then all the characteristics of a discrete Alfv\'en mode.

Finally, at a frequency  of  $\approx 30$ Hz,  we observe excitation mainly of the core region but we notice also significant excitation in the crust and thus this frequency can be identified as  a global oscillation, that is localised mainly in the core and then  forces  the crust to oscillate violently.  

Note that the higher frequencies are located closer to the magnetic axis than the lower ones. This behaviour was already seen in \cite{2009MNRAS.396.1441C} and \cite{2009MNRAS.397.1607C}. Note also that the part of the star that  seems not to be excited corresponds to the closed magnetic field lines. As we already noted in \cite{2009MNRAS.396.1441C}, the closed magnetic lines have a significant smaller amplitude  than the open magnetic field lines and cannot be observed via oscillations or thermal surface phenomena, since they are confined in  the interior of the star. 
 
Concerning the SGR 1900+14, an unique  model  that can fit better the observed frequencies cannot be found because of the paucity of the observations. For this reason, we find that  two models can fit the identified QPOs for the SGR 1900+14: the EOS APR with a mass $M=1.4  M_{\odot}$ , radius $R=11.57$km and a magnetic field strength $B=4.25\times 10^{15}$G, as well as the  EOS WFF with    mass  and  radius respectively $M=1.4\;M_{\odot}$ and $R=10.91$km, and a magnetic field strength $B=4\times 10^{15}$G.   We focus our study on the identification of the  first three frequencies observed: 28Hz, 53Hz and 84Hz.
In the case of APR$_{14}$with a magnetic field strength $B=4.25\times 10^{15}$G, all the three frequencies can be identified as discrete Alfv\'en modes, see upper panel of Fig. \ref{FFT_WFF}. The oscillations involve both crust and core. 
Contrary, in the case of model WFF$_{14}$ (see lower panel of Fig. \ref{FFT_WFF}), the frequencies can be identified as global modes (the 28Hz and the 54Hz) and as crustal mode (the 84Hz). 
In particular,  the 28Hz frequency is identified as the first edge of the second continuum  while the 54Hz is identified as the second edge of the third continuum  (see Fig.  \ref{FFT_WFF}): both those frequencies are global modes, with excitations that involve both the crust and the core. The 84Hz is a crust frequency: its nature is discrete and the oscillations, although initially confined in the crust, expand rapidly in the core. Also in this model, we find a discrete Alfv\'en  mode around 26Hz.

Note that the equation of state used for the crust in all the models that we present in this paper is the NV. As already pointed out in \cite{SKS2007}, the use of the EOS proposed by \cite{2001A&A...380..151D} for the crust does not alter significantly the results. Still since it produces a thinner crust, it creates difficulties in the simulations since it demands finer  grid in the crust in order to evolve properly the perturbations inside it. 

The results present in this section partially agree with the ones found by \cite{2008MNRAS.385.2069L}, where a similar problem of a star with solid crust and fluid core both permeated by a magnetic field is studied in Newtonian theory  as an eigenvalue problem. Without time evolutions only discrete Alfv\'en frequencies could be found and thus no continuous spectrum appears. However, the discrete Alfv\'en frequencies that we found here agree qualitatively with the ones described by Lee. 

It is worth  mentioning that we analysed also  alternative combination of magnetic field  configuration, consisting of both poloidal and toroidal magnetic fields. 
We found that the presence of a toroidal component in the background magnetic field shifts the spectrum towards lower values. This extra parameter complicates significantly the procedure of identifying the magnetar model that will better fit the observed frequencies.

Note that a more accurate treatment of the shear modulus than the one used in equation (\ref{mu}) may change quantitative  the crustal modes. In addition, the presence of a superfluid component  in the inner crust and/or in the core could affect the frequencies significantly,  as discussed in \cite{2009MNRAS.396..894A}.

%---------------------------------------

%---------------------------------------

\section{Conclusions}
%%%%%%

We studied the torsional oscillations of a magnetar in a general relativistic framework, consisting of a relativistic neutron star  with a solid crust and a fluid core. The core and the crust are coupled by the strong magnetic field and this coupling is defined by the appropriate boundary conditions.  We find that the presence of a crust and its coupling with the core partially alters  the earlier  picture presented in \cite{2009MNRAS.396.1441C}.  

In particular, the presence of a crust makes the spectrum denser and thus offers an explanation of the nature of all the observed frequencies. 
%We find also some relation between fundamental frequency and overtones frequencies and 
We can distinguish, using the eigenfunction,  between global modes (i.e. modes for which both the crust and the core oscillate), crust modes and discrete Alfv\'en modes  (i.e. modes that have a discrete nature but that scale with the magnetic field).  
Unlike the paper by \cite{2011MNRAS.410.1036H}, we find that also in the case of intermediate magnetic field ($B<10^{15}$G), the oscillations is not confined to the core but also the crust is excited. We find a  family of modes, the discrete Alfv\'en modes, in the gaps between two contiguous continua (as found by \cite{2011MNRAS.410.1036H}) and, in addition, we can resolve them also inside the continua.

We can identify all the  frequencies observed in SGR 1806-20 and SGR 1900+14 and this allows us to put some constrains on radius, mass, crust thickness and magnetic field strength. In particular,  the frequencies observed in SGR 1806-20 can be fitted uniquely with the APR stellar model with a mass $M=1.4 \; M_{\odot}$, a radius $R=11.57$km, a compactness $M/R=0.178$ and a crust thickness $\Delta r/R=0.099$. The magnetic field strength at the pole is  $B=4\times 10^{15}$G. Contrary, the SGR 1900+14 cannot be strongly constrained because  of the limited information from the observations, i.e just three QPOs were identified.  For this reason, the SGR 1900+14 cannot be reproduced uniquely by a single model, as it has been done  for the SGR 1806-20, since at least two stellar models could reproduce its observed frequencies with a good accuracy: the APR stellar model with a mass $M=1.4 \; M_{\odot}$, a radius $R=11.57$km, a compactness $M/R=0.178$ and a crust thickness $\Delta r/R=0.099$  but with a magnetic field strength  $B=4.25\times 10^{15}$G and  the WFF model with a mass $M=1.4M_{\odot}$, a radius $R=10.91$km, a compactness $M/R=0.189$, crust thickness $\Delta r/R=0.085$ and a magnetic field strength  $B=4\times 10^{15}$G.  Note that, in all the models, the crust is described by the NV equation of state.

\section*{Acknowledgments}%\acknowledgements
This work was supported by the German Science Council (DFG) via SFB/TR7. We thank Erich Gaertig for providing us his recycling routine. We are grateful to  the anonymous referee, to Yuri Levin, to Maarten van Hoven and to Sandro Mereghetti for their useful suggestions and comments.

\bibliographystyle{mn2e}
\bibliography{paper_ac_kk_2010_2}

\end{document}